\documentclass[preprint,12pt]{elsarticle}

\usepackage{graphicx}         % needed for figures
\usepackage{bm}               % for math
\usepackage{amssymb}          % for math
\usepackage{amsmath}          % for aligned

\usepackage[dvipsnames]{xcolor}            % for color
\usepackage[normalem]{ulem}   % for sout

\usepackage{hyperref}
\usepackage[normalem]{ulem}

\begin{document}

%help edit bibliography
\newcommand{\aap}{Astron. Astrophys.}
\newcommand{\aj}{Astronomical Journal}
\newcommand{\apj}{Astrophys. J.}
\newcommand{\apjl}{Astrophys. J. Lett.}
\newcommand{\cqg}{Class.  Quant. Grav.}
\newcommand{\grg}{Gen.  Rel. Grav.}
\newcommand{\jcap}{J. Cosmol. Astropart. Phys.}
\newcommand{\mnras}{Mon. Not. R. Astron. Soc.}
\newcommand{\prd}{Phys. Rev. D}
\newcommand{\prl}{Phys. Rev. Lett}
\newcommand{\plb}{Phys. Lett. B}

\newcommand{\rdm}[1]{{\color{blue} #1}}
\newcommand{\idm}[1]{{\color{orange} \textbf{#1}}}

\newcommand{\reply}[1]{\textcolor{NavyBlue}{\textbf{#1}}}

\begin{frontmatter}

\title{Dwarf galaxies in non-local gravity}

%authors and affiliations
\author[1]{Ivan De Martino\footnote{E-mail: ivan.demartino@usal.es} }
\author[2,3]{Riccardo Della Monica\footnote{E-mail: rdellamonica@usal.es}}
\author[4]{Mahmood Roshan\footnote{E-mail: mroshan@um.ac.ir} }

\address[1]{Departamento de F\'isica Fundamental and IUFFyM, Universidad de Salamanca,Plaza de la Merced, s/n, E-37008 Salamanca, Spain}
\address[2]{CENTRA, Departamento de Física, Instituto Superior Técnico – IST\\
Universidade de Lisboa – UL, Avenida Rovisco Pais 1, 1049-001 Lisboa, Portugal}
\address[3]{Departamento de F\'isica Fundamental, Universidad de Salamanca,\\Plaza de la Merced, s/n, E-37008 Salamanca}
\address[4]{Department of Physics, Faculty of Science, Ferdowsi University of Mashhad, P.O. Box 1436, Mashhad, Iran}

% abstract
% ----------------------------------------------

\begin{abstract}

The nature of dark matter remains one of the most pressing open questions in modern cosmology. Despite extensive experimental efforts, no direct or indirect detection of dark matter particles has been confirmed. This has motivated alternative approaches, including modifications to the underlying theory of gravity. In this work, we investigate the implications of a specific non-local gravity (NLG) theory, which modifies General Relativity by introducing non-local effects that manifest as an effective dark matter component. We analyze the velocity dispersion profiles of eight classical dwarf spheroidal (dSph) galaxies - Carina, Draco, Fornax, Leo I, Leo II, Sculptor, Sextans, and Ursa Minor - to test the predictions of NLG. Using the Jeans equation, we model the kinematics of these galaxies and perform a Bayesian Markov Chain Monte Carlo analysis to constrain the parameters of the NLG kernel chosen for our analysis. {Our results indicate that NLG might successfully reproduce the observed kinematics of dSph galaxies without requiring particle dark matter, providing constraints on the scale-dependent modifications to gravity that are compatible with previous studies in the literature.
{However, a parameter inconsistency remains in the cases of Fornax and Sextans galaxies that requires further attention.}}

\end{abstract}


\end{frontmatter}

\section{Introduction}

The persistent elusiveness of dark matter particles, despite decades of experiments seeking for direct or indirect detection, has led to modified theories of gravity being considered as a potential solution to the dark matter problem \cite{deMartino2020}. One of the primary objectives of the dark matter hypothesis is to explain cosmic structure formation. It is well understood that achieving a viable cosmological structure formation, consistent with cosmic microwave background (CMB) observations, solely through modifications in gravitational physics is extremely challenging. The only modified gravity theory that claims to successfully account for structure formation is the new relativistic theory for Modified Newtonian Dynamics (MOND) theory, as presented in \cite{skordis}. This theory incorporates a scalar field and a vector field, in addition to the metric tensor, to describe gravity. From this perspective, theory is quite complex. Among the various alternative theories proposed to replace dark matter particles, we are particularly interested in a specific non-local gravity (NLG) theory introduced in \cite{Hehl:2008eu}. This theory leverages the analogy between electrodynamics and General Relativity (GR) to incorporate non-local effects into gravitational physics \cite{Hehl:2009es}.  In fact, this theory represents an attempt to generalize GR by incorporating non-local effects, continuing along the pathway established by non-local special relativity \cite{Mashhoon2017}. Although this theory does not introduce any new gravitational fields beyond those in GR, with the metric tensor being the sole field, non-local effects manifest in a complex manner. This complexity has resulted in the full cosmological behavior of theory remaining largely unexplored. On the other hand, the Newtonian limit of theory has been well-studied  \cite{Mashhoon2017}. Consequently, this theory has been extensively applied to astrophysical systems where relativistic effects are negligible, yet it suggests that non-local effects are significant enough to potentially replace dark matter particles. For instance, the rotation curves of spiral galaxies within the framework of NLG have been examined in \cite{Rahvar2014}. The time evolution of disk galaxies using N-body simulations has been investigated in \cite{Roshan:2019xda}. Additionally, dynamical friction, which plays a crucial role in systems with a large fraction of dark matter, has been studied in \cite{Roshan:2021ljs}. For further studies focused on the cosmological aspects of NLG, we refer the reader to \cite{Taba1,Taba2,Taba3}. 

In the case of dwarf galaxies, NLG makes a clear prediction: the smaller the baryonic content of an astrophysical system, the smaller the amount of effective dark matter that NLG predicts. It is worth noting that, as will be detailed in the next section, the non-local aspects of NLG effectively manifest as a dark matter component. In this paper, we will investigate the prediction mentioned above, by focusing on the velocity dispersion observations of eight dwarf spheroidal (dSph) galaxies.  
{The dSph galaxies are known to be dark matter dominated systems and, therefore, they serves as a valuable test-bed for alternative theories of gravity that modify the underlying gravitational field to avoid the introduction of dark matter to explain the kinematics of stars in galaxies.} For example, they have been used to constrain $f(R)$-gravity and Scalar-Tensor-Vector theory both showing a Yukawa-like modification of the gravitational potential in the weak-filed limit \cite{deMartino2023, DeMartino2023_mog}. Other examples of studies investigating modified theories of gravity using kinematic data of dwarf galaxies focused on Degenerate Higher-Order Scalar Tensor (DHOST) theory \cite{Laudato2022a, Laudato2023}, or non-local corrections to the Newtonian potential \cite{Bouche2024}. 

The outline of this paper is as follows: in Section \ref{sec:non_local_gravity}, we summarize the main features of the Poisson equation in the weak field limit of NLG that will serve as the starting point for modelling the velocity dispersion profiles. The latter will be the subject of Section \ref{sec:Jeans_analysis} where we explain all the ingredients necessary to theoretically build the dispersion velocity profile. Then, in Section \ref{sec:Data_and_data_analysis}, we illustrate the data and the data analysis methodology that we will use to constrain NLG. In Section \ref{sec:results}, we expose and discuss our results and, finally, in Section \ref{sec:conclusions} we give our final conclusions.

\section{Gravitational potential in NLG}\label{sec:non_local_gravity}

{Due to the complex nature of the field equations in NLG, no exact solutions have been found so far. Additionally, a modified version of the Friedmann equations has yet to be developed within the NLG framework. However, as expected, the situation is much simpler in the Newtonian limit. In this regime, it has been shown that NLG introduces a Yukawa-like correction to the gravitational potential. More specifically,} the revised version of the Poisson equation in the weak field limit of NLG is presented as \cite{Mashhoon2017}
\begin{equation} \label{eq:Potential_NLG}
 \nabla ^2 \Phi (\mathbf{x})  =  4\pi G[ \rho (\mathbf{x})+\rho_D(\mathbf{x}) ]\,.
\end{equation}
Here, $\rho_D$ represents the effective dark matter density in NLG. It is important to clarify that within NLG, there is no actual dark matter present. Instead, the nonlocal characteristics of gravity manifest as an effective dark matter density in the Poisson equation. This effective density is determined by:
\begin{equation}\label{edm}
\rho_{D}(\mathbf{x})=\int q(|\mathbf{x}-\mathbf{y}|)\rho(\mathbf{y}) d^3 y\,.
\end{equation}
In cases where a specific kernel $q(|\mathbf{x}-\mathbf{y}|)$ is known, the effective dark matter density can be derived solely from the distribution of baryonic matter $\rho$. The choice of kernel is crucial as the nonlocal properties in the weak field limit are directly influenced by it. However, there is no definitive method to determine this kernel, leading to postulations primarily based on observational data such as rotation curves of spiral galaxies \cite{Rahvar2014}. {However, a complementary approach, based on the existence of Noether Symmetries in the system, might be used as criteria to theoretically select the kernel \cite{Dialektopoulos:2018qoe}.}

{The effective dark matter distribution in NLG mirrors the symmetries of the baryonic system. For instance, in an axisymmetric galaxy, the effective dark matter is also axisymmetric. This contrasts with the cold dark matter (CDM) scenario, where the dark matter halo in disk galaxies is typically spherical. Because the effective dark matter is derived through a convolution with the baryonic matter, features such as spiral arms, bars, peanuts, and bulges are directly reflected in the effective dark matter distribution as well \cite{kashfi2024}. This distinct distribution, compared to that of cold dark matter, has important consequences. For example, the radial and vertical evolution of disk galaxies in NLG differs from that in CDM. In particular, the evolution of galactic bars shows significant deviations from the standard CDM case, which could provide observational means to distinguish between NLG and cold dark matter scenarios. Additionally, it has been demonstrated that the effective dark matter in NLG does not suffer from the core-cusp problem observed in galaxies \cite{Roshan2022}}. For further insights into the effective dark matter distribution at galactic scales, we refer the reader to \cite{Roshan2022}. One commonly used kernel at galactic scales is the following \cite{Mashhoon2017}: 
\begin{equation} \label{eq:kernel_NLG}
 q_0(r) = \frac{1}{4 \pi \lambda_0} \frac{1 + \mu_0 r}{r^2} e^{- \mu_0 r}\,,
\end{equation}
where $\mu_0$ and $\lambda_0$ are free parameters. The observations of nearby spiral galaxies and clusters of galaxies imply that $\lambda_0\approx3\pm 2\, \text{kpc}$ and $\mu_0\approx 0.059 \pm 0.028\,\text{kpc}^{-1} $ \cite{Rahvar2014}. On the other hand, the best value of these parameters to fit the rotation curve data of some ultra-diffuse galaxies (UDGs) is $\lambda_0=2.42^{+1.02}_{-0.84}$ kpc and $\mu_0=0.07^{+0.02}_{-0.01}\, \text{kpc}^{-1}$ \cite{Roshan2022}, that is somehow consistent with those obtained from rotation curves of normal galaxies. {In fact, gas-rich UDGs exhibit a baryonic-to-total mass fraction that is significantly higher than that of typical galaxies with similar rotation curves. This implies that UDGs contain less dark matter, which is consistent with the predictions of NLG. However, it is important to highlight the special case of the UDG Dragonfly 44. Unlike others, this galaxy is not rotationally dominated and is believed to be dominated by dark matter \cite{vanDokkum:2019fdc}. Consequently, it is reasonable to expect that NLG may encounter difficulties in explaining the observed velocity dispersion of this galaxy. A thorough investigation of this issue is necessary but falls beyond the scope of the present paper.}

In this paper we study the implications of NLG in dSphs. The baryonic matter density of dSphs is mainly described by the Plummer model \cite{Walker2009d}. Therefore, the baryonic matter density can be written as $\rho_*(r) =  \Upsilon \nu(r)$, where $\Upsilon=M_*/L_V$ is the stellar mass to light ratio, and $\nu(r)$ is given by
\begin{equation}  \label{eq:3.20}
\nu(r)= \frac{3 L_{V}}{4\pi r_{1/2}^3}\left(1+\frac{r^2}{r_{1/2}^2}\right)^{-\frac{5}{2}}\,,
\end{equation}
where $L_V$ is the total luminosity, $r_{1/2}$ is the radius enclosing $0.5\,L_V$. Given that effective dark matter adheres to the symmetries of baryonic matter, we anticipate a spherical distribution for it too. Consequently, using the Newton's shell theorem one may write:
\begin{equation}\label{f1}
\frac{d\Phi}{dr}=\frac{G M(r)}{r^2}+\frac{G M_D(r) }{r^2}\,,
\end{equation}
where $M(r)$ and $M_D(r)$ represent the baryonic and effective dark matter mass within the give radius $r$. $M(r)$ is given by the following analytic expression:
\begin{equation}
M(r)=4\pi \int_0^r \rho_*(y)y^2 dy= \Upsilon L_{V}\left(1+\frac{r^2}{r_{1/2}^2}\right)^{-\frac{3}{2}}\frac{r^3}{r_{1/2}^3}\,.
\end{equation}

\begin{figure}
    \centering
    \includegraphics[width=0.49\textwidth]{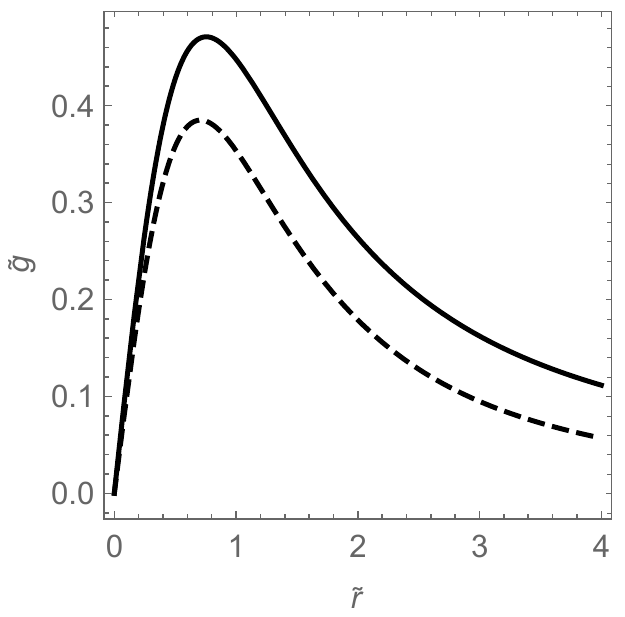}\caption{The dashed curve represents the gravitational field contribution from baryonic matter, while the solid curve illustrates the total gravitational field, incorporating the contribution from effective dark matter, for the case with $\tilde{\mu}_0=0.01$ and $\tilde{\lambda}_0=4$.}
    \label{tildeg}
\end{figure}

On the other hand to obtain $M_D(r)$, let's simplify the effective dark matter density by defining a new variable $\mathbf{z}=\mathbf{x}-\mathbf{y}$. Therefore, $\rho_D$ in equation \eqref{edm} can be rewritten as
\begin{equation}\label{edm2}
\rho_{D}(\mathbf{x})=\int q_0(z)\rho_*(\mathbf{x}-\mathbf{z}) d^3 z\,,
\end{equation}
which takes the following form for the Plummer profile
\begin{equation}\label{edm3}
\rho_D(r)=\frac{3\Upsilon L_V}{2 r_{1/2}^3}\int_0^{\infty}q_0(z) z^2 dz \int_{-1}^{+1}\Big(1+\frac{r^2+z^2- 2 r z \psi}{r_{1/2}^2}\Big)^{-5/2}d \psi\,,
\end{equation}
where $r=|\mathbf{x}|$ and $z=|\mathbf{z}|$, and $\psi=\cos \theta$. The integral over $\psi$ can be simplified as follows \cite{Roshan2022}:
\begin{equation}\label{edm4}
\rho_D(r)=\frac{\Upsilon L_V r_{1/2}^2}{2 r}\int_0^{\infty}q_0(z) z dz \{[r_{1/2}^2+(z-r)^2]^{-3/2}-[r_{1/2}^2+(z+r)^2]^{-3/2}\}\,.
\end{equation}
Finally, for the effective dark matter mass $M_D(r)=4\pi \int_{0}^{r}\rho_D(y)y^2 dy$ we have:
\begin{equation}\label{edm5}
M_D(r)=2\pi \Upsilon L_V \int_0^{\infty}q_0(z) z dz \Big[\frac{r_{1/2}^2+z(z+r)}{\sqrt{r_{1/2}^2+(z+r)^2}}-\frac{r_{1/2}^2+z(z-r)}{\sqrt{r_{1/2}^2+(z-r)^2}}\Big]\,.
\end{equation}
Both integrals in equations \eqref{edm4} and \eqref{edm5} require numerical computation. Now let's define the dimensionless length parameters $\tilde{Q}=Q/r_{1/2}$ (like $\tilde{\lambda}_0=\lambda_0/r_{1/2}$), and $\tilde{\mu}_0=\mu_0 r_{1/2}$, and use the kernel \eqref{eq:kernel_NLG} in order to rewrite the equation \eqref{f1} as follows
\begin{equation}\label{mr1}
\tilde{g}(\tilde{r})=\Big(\frac{\Upsilon L_V G}{r_{1/2}^2}\Big)^{-1}\frac{d\Phi}{dr}=\frac{\tilde{r}}{(1+\tilde{r}^2)^{3/2}}+\frac{1}{2\tilde{\lambda}_0 \tilde{r}^2}\int_0^{\infty} \frac{d\tilde{z} (1+\tilde{\mu}_0 \tilde{z})}{\tilde{z}}e^{-\tilde{\mu}_0 \tilde{z}} f(\tilde{z},\tilde{r})\,,
\end{equation}
where $f(\tilde{z},\tilde{r})$ is defined as
\begin{equation}
f(\tilde{z},\tilde{r})=\frac{1+\tilde{z}(\tilde{z}+\tilde{r})}{\sqrt{1+(\tilde{z}+\tilde{r})^2}}-\frac{1+\tilde{z}(\tilde{z}-\tilde{r})}{\sqrt{1+(\tilde{z}-\tilde{r})^2}}\,.
\end{equation}
The dimensionless gravitational field $\tilde{g}$ is depicted in Figure \ref{tildeg} for selected values of $\tilde{\mu}_0=0.01$ and $\tilde{\lambda}_0=4$. It is clear that non-local corrections enhance the gravitational force. The gravitational field \eqref{mr1} will then be used to compute stars' velocity dispersion profile in dwarf galaxies.

\section{The Jeans Analysis}\label{sec:Jeans_analysis}

{The standard non-relativistic fluid equations remain valid in the Newtonian limit of NLG. The only modification is in the Poisson equation, which now includes a new source term, $\rho_D$, on the right-hand side. Mathematically, this modified equation is identical in form to the standard one. Therefore, the Jeans equations remain unchanged in the Newtonian limit of NLG. Since we are studying the observations of the velocity dispersion of dwarf galaxies, it is necessary to derive an analytic expression for it within the framework of NLG in order to fit the observational data.} The velocity dispersion can be modelled by solving the Jeans equation which, for a static spherical system and in the spherical coordinates system, is written as  \cite{Binney2008} 
\begin{equation}  \label{eq:3.19}
\frac{1}{\rho_*(r)}\frac{d}{dr}[\rho_*(r) \bar{v_r^2}(r)]+2\frac{\beta(r)\bar{v_r^2}(r)}{r}=- \frac{d \Phi(r)}{d r} ,
\end{equation}
where we have introduced the anisotropy parameter
\begin{equation}
\beta(r)=1-\frac{\bar{v_{\theta}^2}+\bar{v_{\phi}^2}}{2\bar{v_r^2}}\,,
\end{equation}
{and the quantities $\bar{v_i^2}$ ($i=r,\,\theta,\,\phi$) represent the averaged squared components of the stellar velocity in the radial and tangential directions}. {Generally speaking, the anisotropy parameter would depend on the distance from the centre of the galaxy \cite{Osipkov1979, Merritt1985}. However, it can only be inferred through dynamical mass modeling based on photometric data but such a modeling typically requires assuming a specific form for the dark matter halo \cite{Baes2007}. Since, in NLG, there is no dark matter halo, so the situation may be different, and we choose to set $\beta(r) = {\rm const}$. Under this assumption, the Jeans equation becomes significantly simpler and the solution to equation \eqref{eq:3.19} can be expressed as follows \cite{Lokas2003}}:
 \begin{equation}
 	\rho_*(r)\bar{v_r^2}(r)=r^{-2{  \beta}}\int_{r}^{\infty} \frac{d\Phi(x)}{dx}\rho_*(x)x^{2{  \beta}} \,dx \,.
 \end{equation}
 The stellar density is $\rho_*(r)=\Upsilon\nu(r)$ where $\nu(r)$ is given by \eqref{eq:3.20}.
We then project the solution of the equation \eqref{eq:3.19} along the line-of-sight (LOS) obtaining
\begin{equation}\label{eq:sigmalos}
\sigma_{\rm LOS}^2(R)=\frac{2}{\Sigma(R)}\int_{R}^{\infty}\biggl (1-\beta\frac{R^2}{r^2}\biggr ) \
              \frac{\rho_*(r) \bar{v_r^2}r}{\sqrt{r^2-R^2}}dr,
\end{equation}
where $R$ is the projected radius, $\sigma_{\rm LOS}(R)$ is the LOS velocity dispersion that we can directly compare to the data and, $\Sigma(R)$ is the surface mass stellar density that can be derived by the three-dimensional mass density profile 
once the latter is projected, resulting in
\begin{equation}\label{eq:surfacePlummer}
    \Sigma(R) = \frac{\Upsilon L_V}{\pi r_{1/2}^2}\left(1+\frac{R^2}{r_{1/2}^2}\right)^{-2}\,.
\end{equation}

At this stage, we have assembled all the necessary components. theoretical velocity dispersion profile of the stars depends on the parameters $\lambda_0$, $\mu_0$, $\beta$, and $\Upsilon$, which will be constrained by fitting equation \eqref{eq:sigmalos} to the data. 

\section{Data and data analysis}\label{sec:Data_and_data_analysis}

In our analysis, by solving the Jeans equation illustrated in Sect. \ref{sec:Jeans_analysis}, we predict theoretical velocity dispersion profile projected along the line of sight by solving Eq. \eqref{eq:sigmalos} and fit it to the observed LOS velocity dispersion profiles of eight dSphs: Carina, Fornax, Sculptor, Sextans, Draco, Leo I, Leo II, and Ursa Minor, whose physical properties are detailed in Table \ref{tab:1}. Through this process, we aim to derive the values of the parameters of the non-local gravity parameters $\lambda_0$ and $\mu_0$, the velocity anisotropy parameter $\beta$, and the stellar mass-to-light ratio $\Upsilon$ for each galaxy, along with their corresponding uncertainties using an MCMC analysis. In Sect. \ref{sec:data} and \ref{sec:mcmc} we will describe the data set we will use and the statistical analysis we will carry out, respectively.

\subsection{Data}\label{sec:data}

\begin{table}
		\begin{center}
			\resizebox{12cm}{!}{
				\setlength{\tabcolsep}{12pt}
                \renewcommand{\arraystretch}{1.5}
				\begin{tabular}{|lcccc|}
					\hline
					Galaxy & $\log(L_{\rm V})$  & $r_{1/2}$ & $\Upsilon$  &Ref. \\
					  &($L_{\odot}$)      & (pc)     &  ($\frac{M_{\odot}}{L_{\odot}}$)  &  \\
					(1)         & (2)      & (3)  & (4) & (5)   \\
					\hline
					\textbf{Carina} &  5.57$\pm$0.20 & 273$\pm$45 & $3.4\pm2.9$  & \cite{Pietrzynski2009,Irwin1995,Walker2009c,Fritz2018,deMartino2023}  \\[0.1cm]
					\textbf{Draco} & 5.45$\pm$0.08 & 244$\pm$9 & $11.1\pm4.7$ &  \cite{Walker2009c,Bonanos2004,Martin2008,Walker2007,Fritz2018,deMartino2023}\\[0.1cm]
					\textbf{Fornax} & 7.31$\pm$0.12 & 792$\pm$58&  $7.1\pm6.0$ & \cite{Pietrzynski2009,Irwin1995,Walker2009c,Fritz2018,deMartino2023} \\[0.1cm]
					\textbf{Leo I} & 6.74$\pm$0.12 & 298$\pm$29& $8.8\pm5.6$&  \cite{Irwin1995,Walker2009c,Bellazzini2004,Mateo2008,Fritz2018,deMartino2023}  \\[0.1cm]
					\textbf{Leo II} &  5.87$\pm$0.12 & 219$\pm$52 & $0.4\pm0.4$  & \cite{Irwin1995,Walker2009c,Bellazzini2005,Koch2007,Fritz2018,deMartino2023}\\[0.1cm]
					\textbf{Sculptor} & 6.36$\pm$0.20 & 311$\pm$46 & $3.6\pm2.0$ & \cite{Irwin1995,Walker2009c,Pietrzynski2008,Fritz2018,deMartino2023} \\[0.1cm]
					\textbf{Sextans} & 5.64$\pm$0.20 & 748$\pm$66 & $8.5\pm3.3$ & \cite{Irwin1995,Walker2009c, Lee2009,Fritz2018,deMartino2023}\\[0.1cm]
					\textbf{Ursa Minor} &  5.45$\pm$0.20 & 398$\pm$44 & $1.2\pm1.3$ &\cite{Irwin1995,Walker2009c,Carrera2002,Walker2009b,Fritz2018,deMartino2023}\\[0.1cm]
					\hline
				\end{tabular}
			}
		\end{center}
	\caption{Observational properties of the eight dSphs {analysed in this work}. Columns (2): total $V$-band luminosity; Column (3): half-light radius; Column (4): the stellar mass-to-light ratio estimated by \cite{deMartino2023} using stellar population synthesis models in \cite{Bell2001}; and Column (5): references from which data were extracted.}
    \label{tab:1}
\end{table}

The spectroscopic data sets for Carina, Fornax, Sculptor, and Sextans were acquired using the Michigan/MIKE Fiber Spectrograph \cite{Walker2007, Walker2009a, Walker2009b, Walker2009c, Walker2009d}, while data sets for Draco, Leo I, Leo II, and Ursa Minor were obtained with the Hectochelle fiber spectrograph at the MMT \cite{Mateo2008}. Additionally, luminosity values in the V-band, stellar mass-to-light ratio, and half-light radius for each galaxy are sourced from \cite{Pietrzynski2009,Irwin1995,Walker2009c,Bonanos2004,Martin2008,Walker2007,Bellazzini2004,Mateo2008,Bellazzini2005,Koch2007,Pietrzynski2008,Lee2009,Carrera2002,Walker2009b,Fritz2018} and summarized in Table \ref{tab:1}. 

To determine the velocity dispersion profile, one crucial step is to identify stars within the dSph galaxy. In \cite{Walker2009b}, authors built a membership probability to assign each observed star to the corresponding dSph galaxy using an iterative expectation maximization technique. The parameters that are taken into account in the procedure are the star's position, the magnesium index, and LOS velocity. In \cite{Walker2007, Walker2009a, Walker2009c}, the velocity dispersion profiles were derived by incorporating stars with a membership probability exceeding 95\%. These stars were then grouped into radial circular annuli, each containing an equal number of stars and, finally, the overall transverse motion of the dSph was subsequently subtracted from the analysis.

In a broader context, the overall mass-to-light ratio of a dSph typically hinges on the mass of the dark matter halo. However, in non-local gravity, dark matter is absent, making the mass-to-light ratio required to fit kinematic data sets coincide with the stellar mass-to-light ratio. This estimation can be obtained using the stellar population synthesis models in \cite{Bell2001}. While retaining $\Upsilon$ as a free parameter, as suggested by \cite{deMartino2023, DeMartino2023_mog}, we will assign a Gaussian prior on it according to the averaged values of $\Upsilon$ presented in Table \ref{tab:1}.

\subsection{Methodology}\label{sec:mcmc}

Since we aim to constrain the non-local gravity parameters, namely $\lambda_0$ and $\mu_0$, we design a statistical procedure in which
our theoretical model is the projected velocity dispersion profile in non-local gravity, given by Eq. \eqref{eq:sigmalos} and hereby labelled as $\sigma_{\mathrm{los,\, th}}(r)$, and the latter is fitted to the projected velocity dispersion profile data sets measured by \cite{Walker2009d}  and, hereby, labelled as $\sigma_{\mathrm{los,\, obs}}(r)$. The  parameter space is explored by employing the MCMC algorithm \texttt{emcee} \cite{emcee}
to provide an estimation of the best-fit values and their corresponding uncertainties for the four free parameters: $\bm{\theta} =$ \{$\lambda_0$, $\mu_0$, $\beta$, $\Upsilon$\}. Moreover, we set a uniform prior distribution on $\log[\lambda_0({\rm kpc})]\in [-2;4]$, $\log[\mu_0({\rm kpc}^{-1})]  \in [-8;8]$, and $\beta\in [-20, 1)$. Finally, for each dSph, we set a Gaussian prior on the stellar mass-to-light ratio $\Upsilon$, with mean value and dispersion set according to Table \ref{tab:1} (those values are taken from Column (13) of Table 1 in \cite{deMartino2023}). Finally, the posterior probability  distribution is given by the following likelihood function
    \begin{align}
    	-2\log \mathcal{L}(\bm{\theta}|\textrm{ data}) \propto& \sum_i\biggl[\frac{\sigma_{\mathrm{los,\, th}}(\bm{\theta},\, R_{p,i})-\sigma_{\mathrm{los,\, obs}}(R_{p,i})}{\Delta\sigma_{\mathrm{los,\, obs}}(R_{p,i})}\biggr]^2\,,
    	\label{eq:likelihood}
    \end{align}
where $\Delta\sigma_{\mathrm{los,\, obs}}(r_i)$ indicates the observational uncertainties on the projected velocity dispersion profile data sets $\sigma_{\mathrm{los,\, obs}}(r)$. Finally, for each galaxy we run 12 chains, and we consider they have reached the convergence when the length of each chain is 100 times longer than the autocorrelation time and the latter changes by less than 1\% (for more details we refer to Sec. 3 of \cite{deMartino2022}).

\begin{figure}
    \centering
    \includegraphics[width=0.49\textwidth]{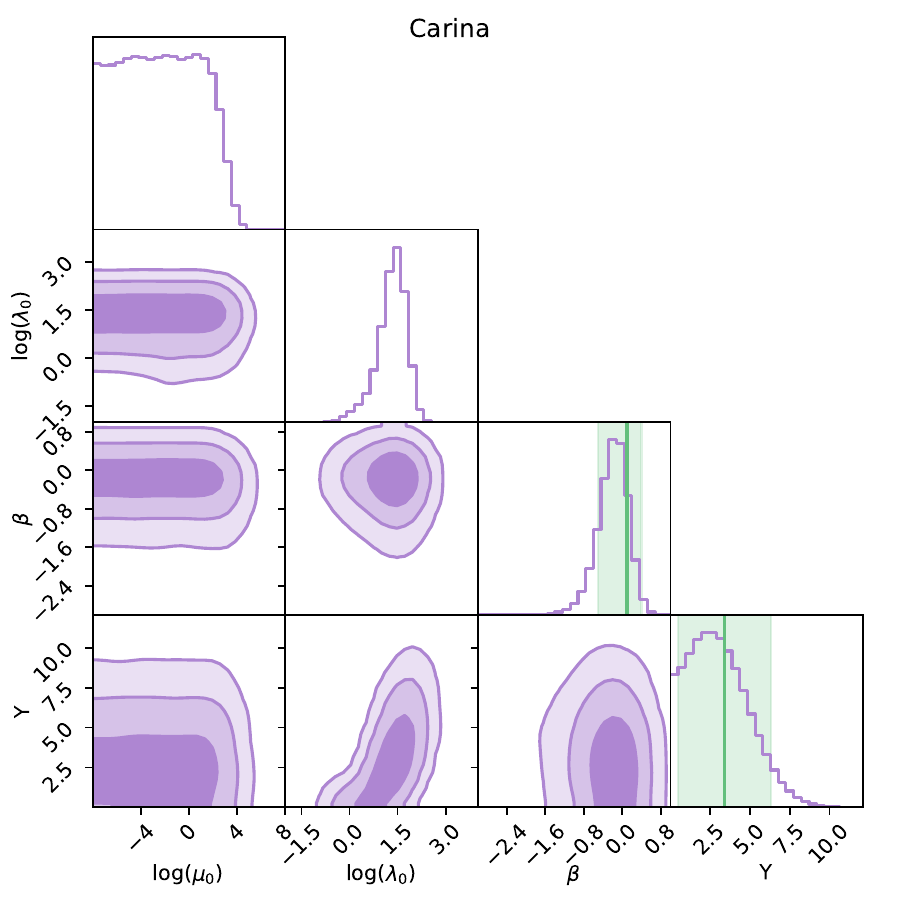}
    \includegraphics[width=0.49\textwidth]{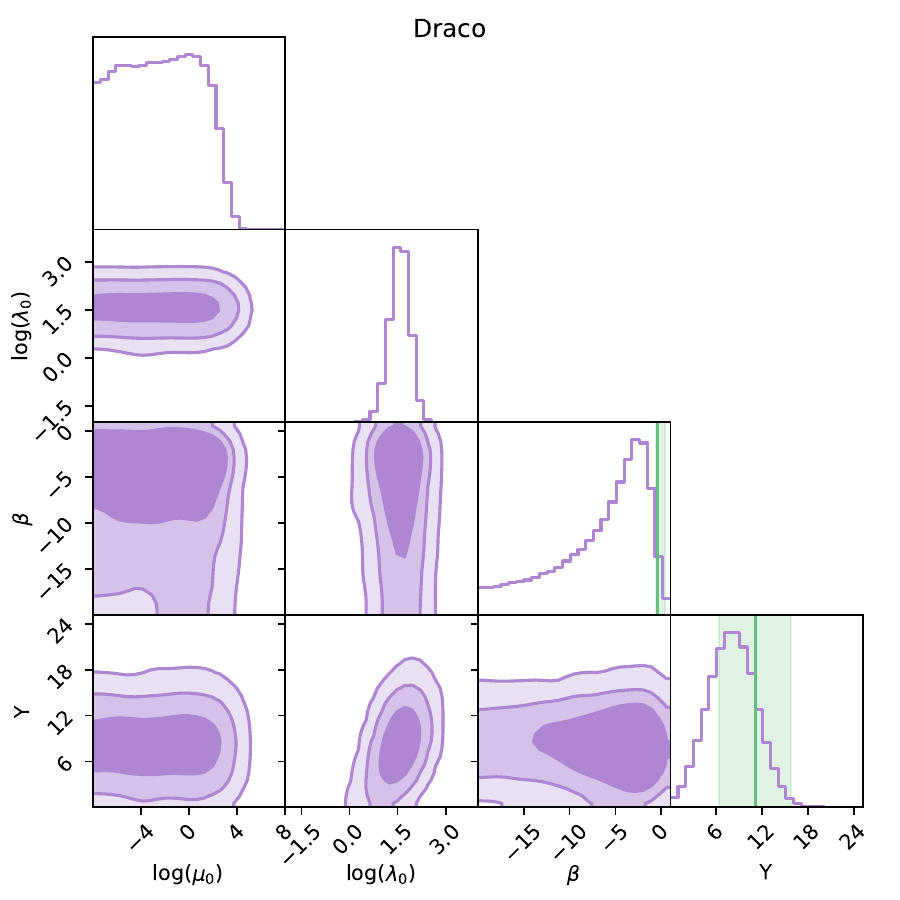}
    \includegraphics[width=0.49\textwidth]{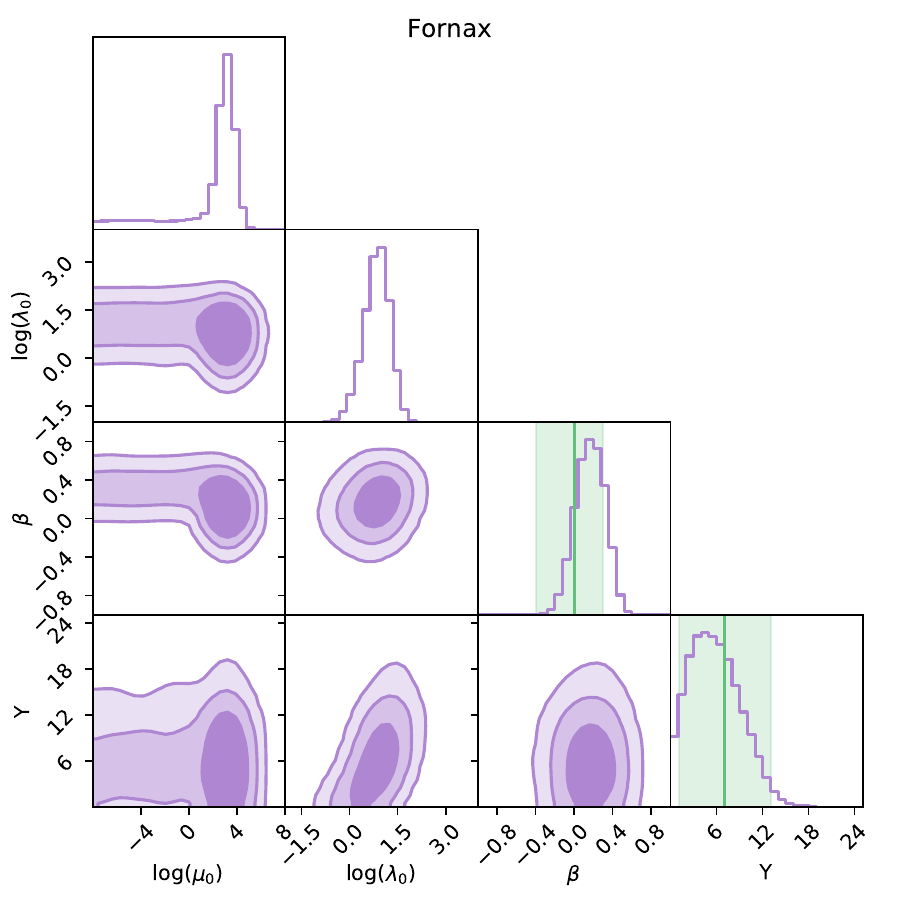}
    \includegraphics[width=0.49\textwidth]{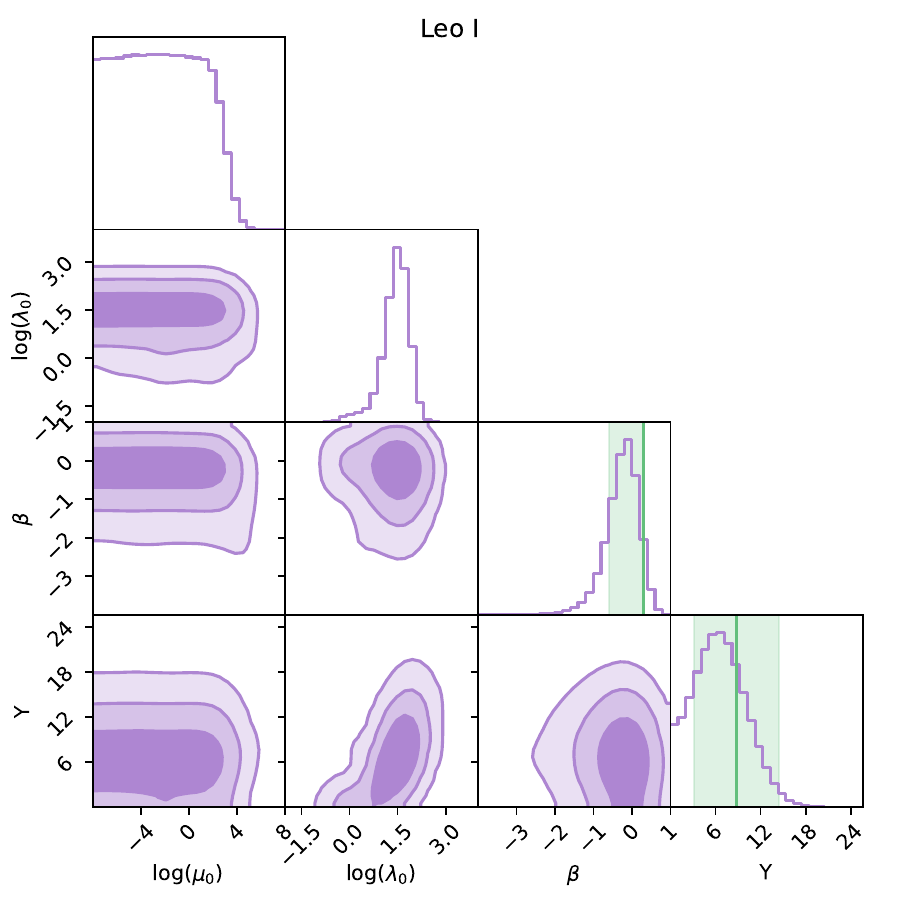}
    \caption{The figure depicts the posterior distributions of the parameters $\bm{\theta} = \{\log\mu_0, \log \lambda, \beta, \Upsilon\}$ for the dSph galaxies Carina Draco, Fornax, and Leo I (as indicated by the titles on top of each corner). The purple-shaded areas with decreasing darkness show the 68\%, 95\%, and 99\% confidence regions, respectively.  The green shaded areas correspond to the values of the velocity anisotropy parameter reported in \cite{Walker2009d},
and the expected values of $\Upsilon$ listed in Table \ref{tab:1}.}
    \label{fig:corner_plots1}
\end{figure}

\begin{figure}
    \centering
    \includegraphics[width=0.49\textwidth]{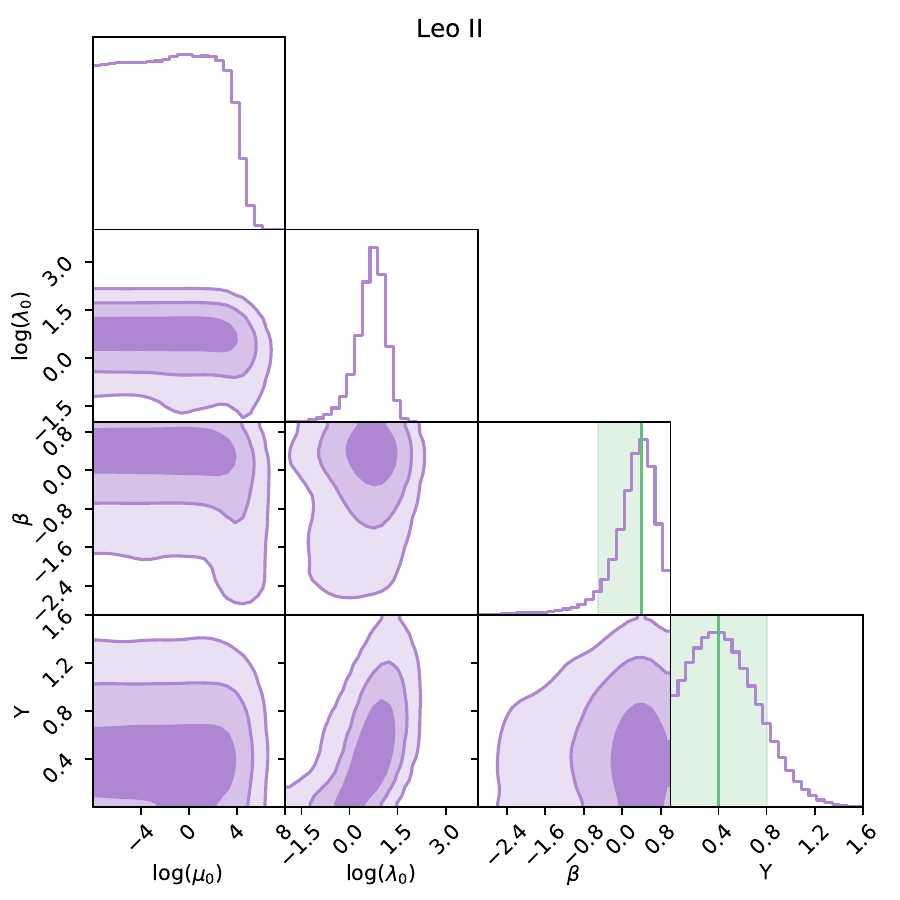}
    \includegraphics[width=0.49\textwidth]{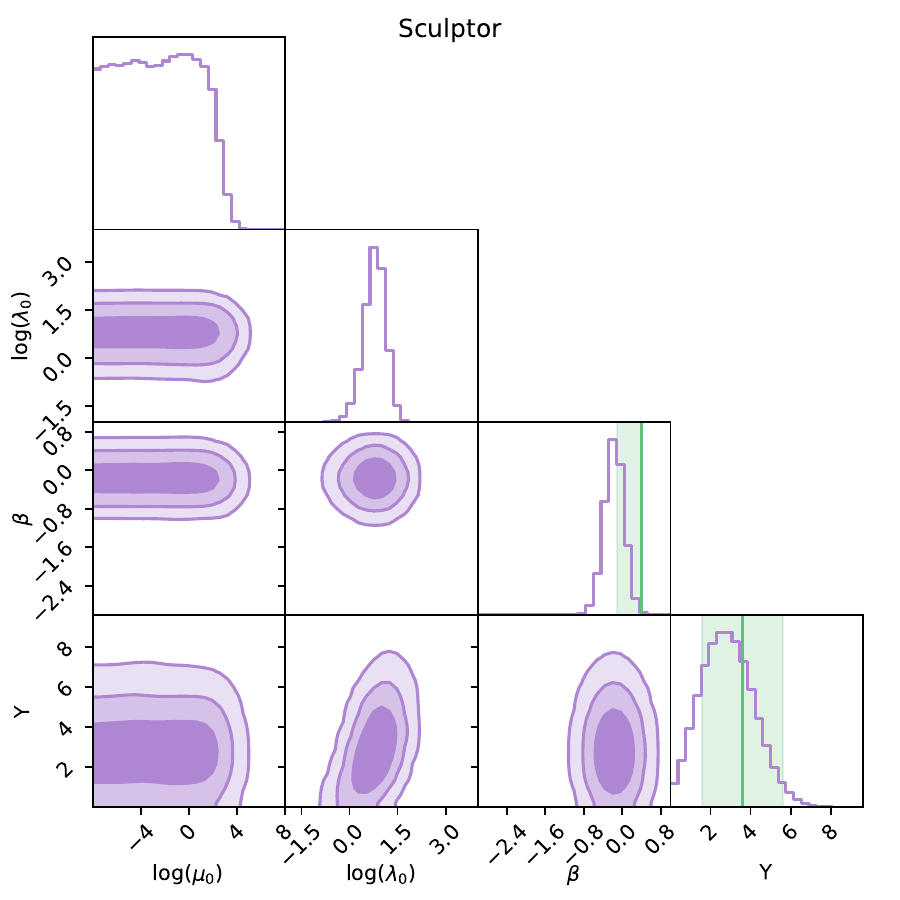}
    \includegraphics[width=0.49\textwidth]{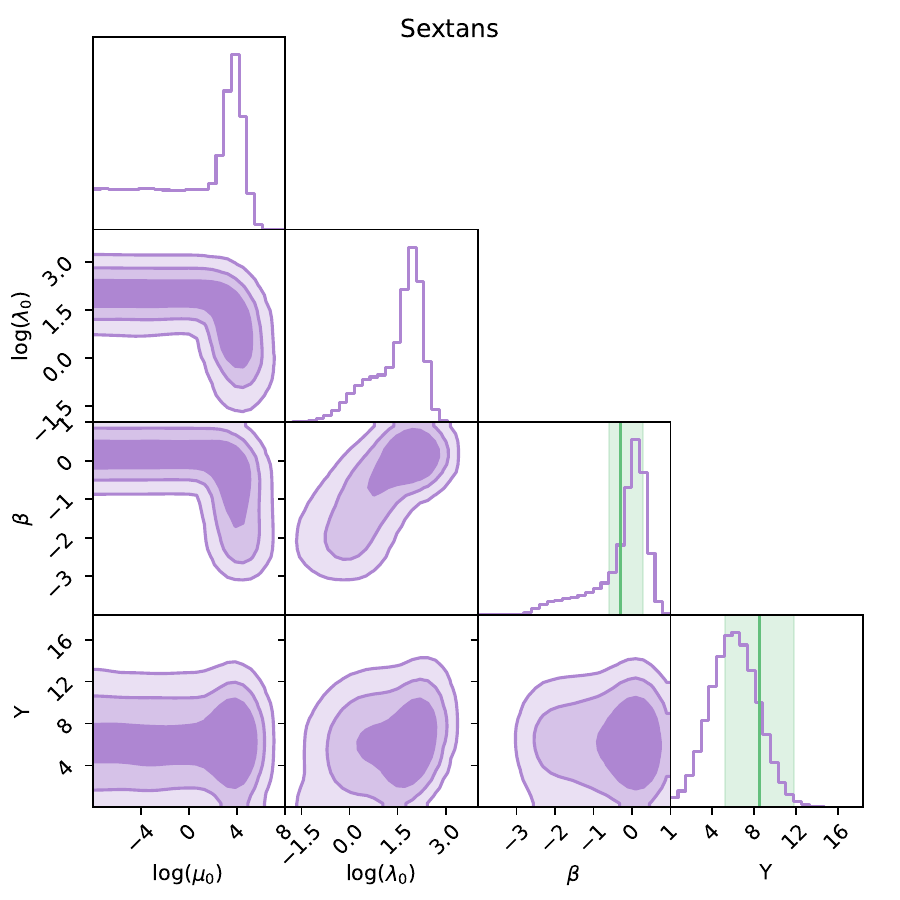}
    \includegraphics[width=0.49\textwidth]{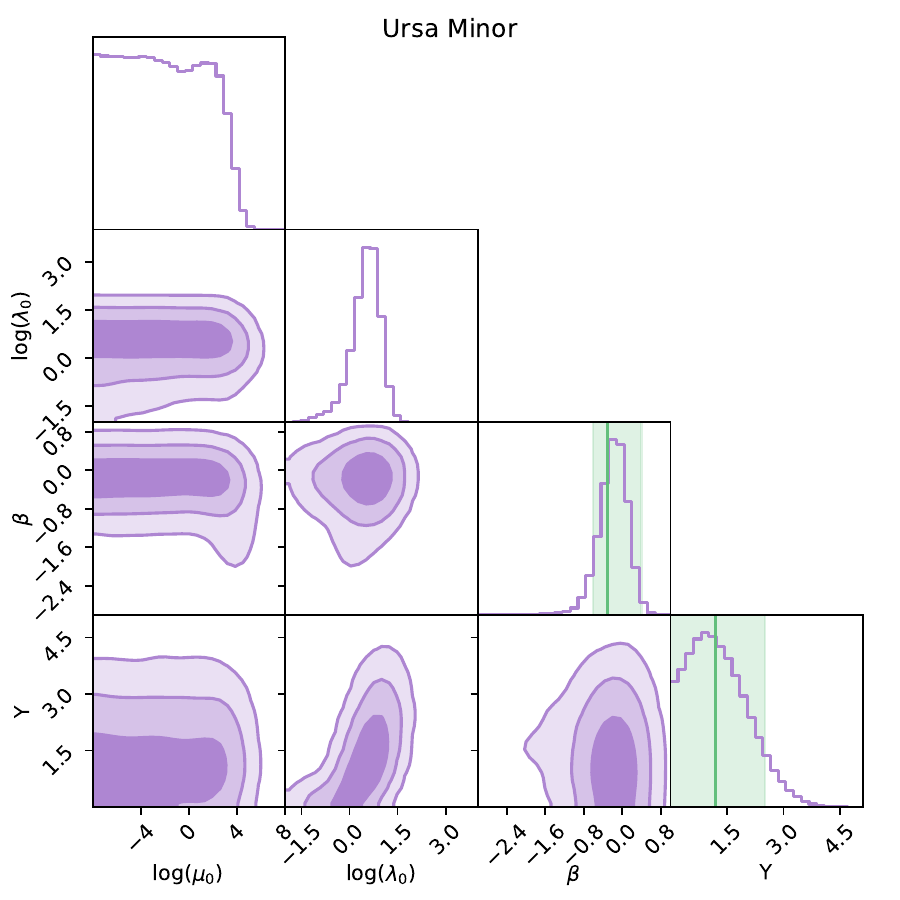}
    \caption{The same of figure \ref{fig:corner_plots1} but for the dSph galaxies Leo II, Sculptor, Sextans and Ursa Minor.}
    \label{fig:corner_plots2}
\end{figure}

\begin{table}
		\begin{center}
			\resizebox{\columnwidth}{!}{
				\setlength{\tabcolsep}{25pt}
				\begin{tabular}{|lcccc|}
					\hline
					Galaxy & $\log\mu_0$ &  $\log\lambda_0$ &$\beta$  &  $\Upsilon$  \\
                                &   (kpc$^{-1}$)       & (kpc)  & & \\
					(1)         & (2)      & (3)  & (4) & (5)   \\
					\hline
        &          &   & & \\[-0.3cm]
					\textbf{Carina} & $\lesssim 2.50$ & $1.3_{-0.34}^{+0.35}$ & $-0.2\pm0.26$ & $3.0\pm1.9$ \\[0.1cm]
					\textbf{Draco} & $\lesssim 2.23$ & $1.56\pm0.18$ & $-6.8_{-6.1}^{+4.8}$ & $8.1_{-3.0}^{+2.9}$ \\[0.1cm]
					\textbf{Fornax} & $3.0_{-4.0}^{+0.08}$ & $0.84\pm0.31$ & $0.16_{-0.13}^{+0.14}$ & $5.9_{-3.3}^{+3.2}$ \\[0.1cm]
					\textbf{Leo I} &  $\lesssim 3.29$ & $0.85\pm0.31$ & $0.17_{-0.13}^{+0.14}$ & $5.9_{-3.3}^{+3.2}$  \\[0.1cm]
					\textbf{Leo II} & $\lesssim 3.90$ & $0.62_{-0.36}^{+0.38}$ & $0.17_{-0.32}^{+0.44}$ & $0.46\pm0.29$ \\[0.1cm]
					\textbf{Sculptor} & $\lesssim 2.10$ & $0.78\pm0.21$ & $-0.17\pm0.12$ & $2.9\pm1.3$ \\[0.1cm]
					\textbf{Sextans} & $1.5_{-6.5}^{+1.9}$ & $1.3_{-1.1}^{+0.8}$ & $-0.5_{-1.0}^{+0.7}$ & $6.1\pm2.2$ \\[0.1cm]
					\textbf{Ursa Minor} & $\lesssim 3.00$ & $0.49_{-0.34}^{+0.35}$ & $-0.19_{-0.22}^{+0.23}$ & $1.26_{-0.81}^{+0.8}$ \\[0.1cm]
					\hline
                    {\bf All galaxies} & $\lesssim3.2$ & $9.7\pm1.9$ & - & - \\[0.1cm]
					\hline
				\end{tabular} 
			}
		\end{center}
	\caption{The Table reports the median and the 68\% confidence intervals of the posterior distribution of the parameters  $\bm{\theta} = \{\log\mu_0, \log \lambda_0, \beta, \Upsilon\}$ for all the dSphs. The last row shows the results of a joint analysis of all eight galaxies. In this case, the upper bound on $\log\log\mu_0$ is set at 99.7\% confidence.}\label{tab:2}
	\end{table}

\section{Results and discussions} \label{sec:results}

We predicted theoretical LOS velocity dispersion profiles using the Jeans analysis explained in Section \ref{sec:Jeans_analysis}, to fit the observational data sets of eight dSph galaxies, namely Carina, Draco, Fornax, Leo I, Leo II, Sculptor, Sextans, and Ursa Minor.  We carried out a MCMC analysis to predict the posterior distribution of the four-dimensional parameter space $\bm{\theta} = \{\log\mu_0, \log \lambda_0, \beta, \Upsilon\}$ of each dSph galaxy in our dataset, thus we estimated the median and the 68\% confidence intervals of the posterior distribution of the parameters $\bm{\theta}$, and we report them in Table \ref{tab:2}. 

Figures \ref{fig:corner_plots1} and \ref{fig:corner_plots2} show, as purple-shaded areas, the 68\%, 95\%, and 99\% confidence regions with decreasing darkness. At the top of each column, we present the one-dimensional marginalized posterior distribution of the corresponding parameter. The green-shaded areas indicate the best-fit values and the $1\sigma$ uncertainties of the velocity anisotropy parameter in \cite{Walker2009d}, as well as the expected values of $\Upsilon$ which is listed in Table \ref{tab:1}.  The stellar mass-to-light ratios align with expectations from the stellar population synthesis model, as a Gaussian prior is set on it. Additionally, the anisotropy parameter $\beta$ consistently falls within the 68\% confidence interval of the value estimated in the standard cold dark matter model \cite{Walker2009c}, except for the Draco dwarf galaxy, where agreement is reached only at the 95\% confidence level. Hence, we can argue that the kinematic structure of dwarf galaxies predicted in NLG aligns to that expected in the cold dark matter, {\em i.e.}, NLG does not introduce radial or tangential biases relative to cold dark matter. 

The rotation curves of  dwarf galaxies in the LITTLE THINGS catalog have been fitted in the NLG adopting $\lambda_0=3.08\pm 1.64$ kpc and $\mu_0=0.059\pm 0.028\, \text{kpc}^{-1}$ \cite{Haghighi:2016jfw}. More recently, the rotation curves of three UDGs, namely, AGC 114905, 242019 and 219533, have been studied leading to the following best the values of NLG parameters $\lambda_0=2.42^{+1.02}_{-0.84}$ kpc and $\mu_0=0.07^{+0.02}_{-0.01}\, \text{kpc}^{-1}$ \cite{Roshan2022}. Our analysis always constrains the parameter $\lambda_0$ while $\mu_0$ results to be almost unconstrained, in fact only an upper limit is listed in the Table \ref{tab:2}, except for the two dSph galaxies Fornax and Sextans. 

\begin{figure}
    \centering
    \includegraphics[width=\textwidth]{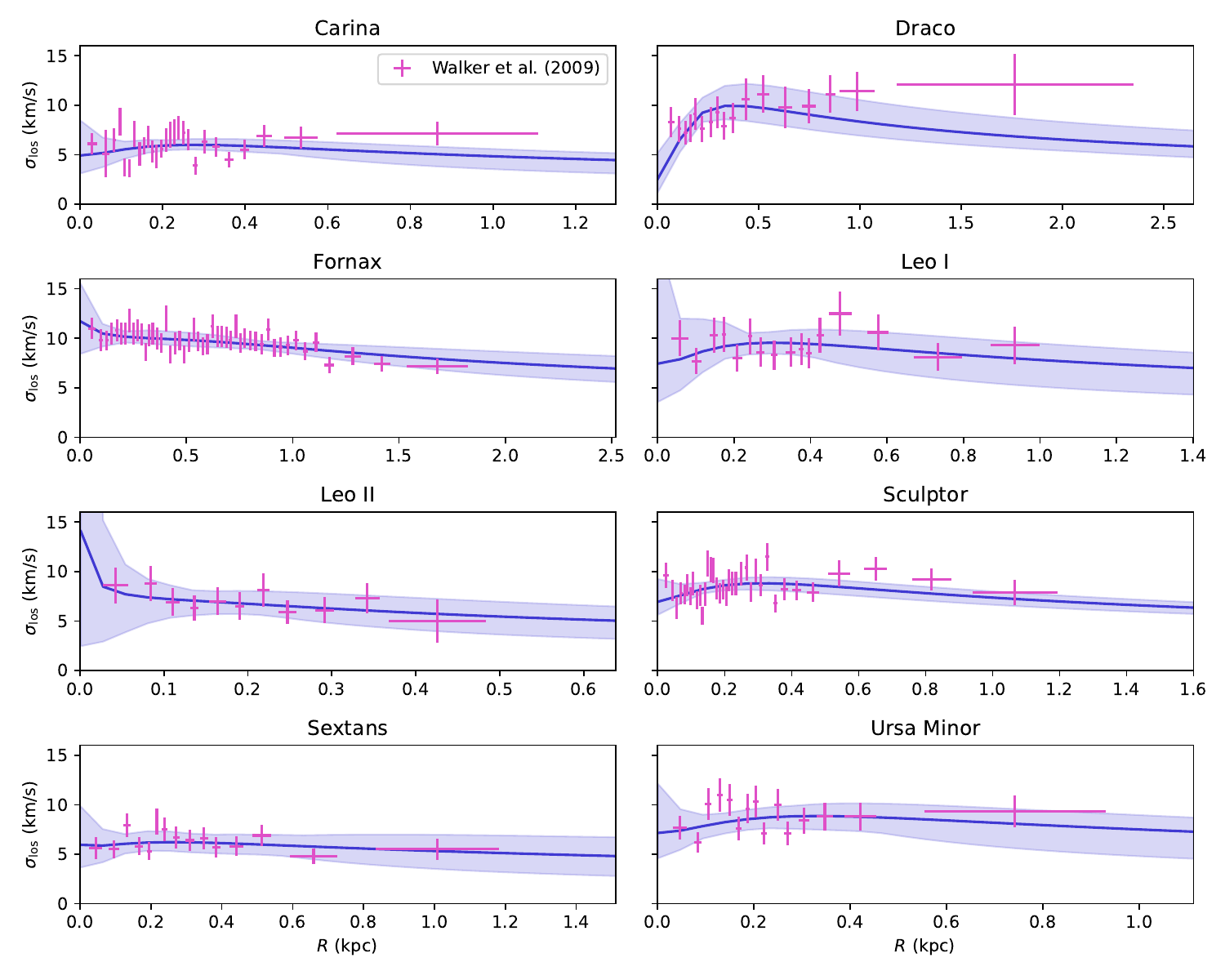}
    \caption{The figure presents the radial profiles of the LOS velocity dispersions for the eight dSphs listed in Table \ref{tab:1}. The purple dots with error bars represent the observational data from \cite{Walker2009c}. The blue solid lines depict the predictions based on NLG using the best-fit parameters $\bm{\theta} = \{\log\mu_0, \log \lambda_0, \beta, \Upsilon\}$ from Table \ref{tab:2}, while the blue-shaded areas indicate the corresponding 68\% confidence interval.}
    %\idm{R as upper case since it is the projected radius.}}
    \label{fig:profiles}
\end{figure}

The Figure \ref{fig:profiles} illustrates the effectiveness of NLG in accurately reproducing the observed LOS velocity dispersion profiles. For each dSph galaxy in Table \ref{tab:1}, the purple circles with error bars represent the observational estimation of the velocity dispersion along the line of sight from \cite{Walker2009c}. The blue solid lines depict the NLG-predicted LOS velocity dispersion profiles based on the best-fit parameters $\bm{\theta} = \{\log\mu_0, \log \lambda, \beta, \Upsilon\}$ listed in Table \ref{tab:2}. The blue-shaded areas indicate the 68\% confidence interval, derived via Monte Carlo sampling of the one-dimensional posterior distributions shown in Figures \ref{fig:corner_plots1} and \ref{fig:corner_plots2}.

\begin{figure}[ht!]
    \centering
    \includegraphics[width=0.6\linewidth]{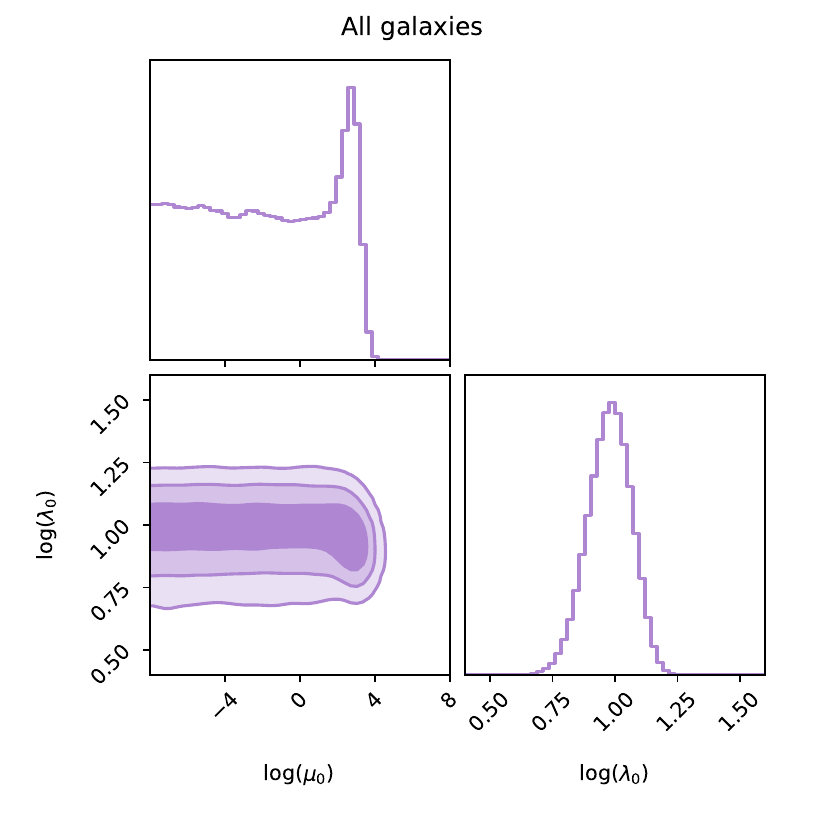}\label{fig: mcmc_all}
    \caption{The figure depicts the results of a joint analysis of all eight galaxies. In this case, the upper bound on $\log\mu_0$ is set at 99.7\% confidence.}
\end{figure}

{Finally, to simplify the comparison with previous analyses, we have repeated our MCMC analysis to the joint set of all eight galaxies in our study, \emph{i.e.} now considering an 18-dimensional parameter space (2 NLG parameters + 2 intrinsic parameters for each galaxy). Once convergence is achieved, we marginalized on all the intrinsic parameters (being all other individual-galaxy parameters bound and compatible to the results with the single-galaxy approach) and focused here on the $\mu_0-\lambda_0$ slice of the parameter space. The results are shown below in the Figure \ref{fig: mcmc_all}.}
{The joint analysis results only in an upper limit on $\log(\mu_0/\textrm{kpc}^{-1})\lesssim3.2$ (at 99.7\% confidence) and in a bounded posterior only for $\lambda_0 = 9.7\pm1.9$ kpc. In the case of $\lambda_0$, this parameter results to be compatible at the 68\% of confidence level with the value used to analyse rotation curves from the LITTLE THINGS catalog, and at the 95\% of confidence level with the value obtained from the analysis of UDGs. Conversely, the parameter $\mu_0$  can only be upper bounded and it results to be poorly constrained and largely overlapping with the previous constraints based on dwarf and UDGs. }

\section{Conclusions}\label{sec:conclusions}

NLG gravity modifies the Poisson equation for the
Newtonian gravitational potential with an additional matter density term $\rho_D(r)$ which brings nonlocal effects to the gravitational interaction. This extra matter density can act as an effective dark matter component. However, the analytic calculation of $\rho_D(r)$ is challenging, as it involves a convolution between the density of baryonic matter and an empirical kernel. Such quantities can be set to the Plummer profile, given in Equation \eqref{eq:3.20}, and a kernel commonly used at the galactic scales in Equation \eqref{eq:kernel_NLG}, respectively. These choices allowed us to solve the Poisson equation and carry out a Jeans analysis of eight dSph galaxies whose observational features are listed in Table \ref{tab:1}. Finally, we performed a MCMC analysis to estimate the best fit values of the NLG parameters $\lambda_0$ and $\mu_0$. The results of our analysis are listed in Table \ref{tab:2} and shown in Figures \ref{fig:corner_plots1}, \ref{fig:corner_plots2} and \ref{fig:profiles}.

Our results are compatible with previous findings in NLG and add another piece to the picture since the mass-to-light ratio adopted as Gaussian prior in our analysis, and subsequently correctly recovered by our Bayesian analysis, is based on the stellar population synthesis models and, therefore, it does not lead to any inconsistency with observations as it could happen if $\Upsilon$ were completely free to vary, as found by previous  analysis. Finally, we found that NLG is capable of describing the velocity dispersion of dSph and recovers values of the anisotropy parameter compatible at the 68\% level with those from the CDM model. Therefore, the kinematic structure of the galaxies is similar in both the NLG and CDM which confirms the capability of NLG in mimicking the effect of the presence of a dark matter component. {Nevertheless, in the case of Fornax and Sextans galaxies, data were capable to bound the $\mu_0$ parameter whose best fit value is only marginally compatible ($\sim 1.5 \sigma$) with other constraints in the literature. Whether or not this may be interpreted as a sign of an observational tension in the NLG gravity must be accurately studied when better data will be available.}

\textit{Acknowledgements}  ---  IDM and RDM acknowledge financial support from the grant PID2021-122938NB-I00 funded by MCIN/AEI/10.13039/501100011033. Moreover, IDM also acknowledges support from the grant SA097P24 funded by Junta de Castilla y Le\'on and by "ERDF A way of making Europe". RDM also acknowledges support from Consejeria de Educación de la Junta de Castilla y León, the European Social Fund + and the Project ``GravNewFields'' funded under the ERC-Portugal program. The work of MR is supported by the Ferdowsi University of Mashhad.

\bibliographystyle{elsarticle-num}
\bibliography{biblio}

\end{document}